\documentclass{article}
\usepackage[T1]{fontenc} % add special characters (e.g., umlaute)
\usepackage[utf8]{inputenc} % set utf-8 as default input encoding
\usepackage{ismir,amsmath,cite,url}
\usepackage{graphicx}
\usepackage{color}
\usepackage{multirow}
\usepackage{subfig}
\usepackage{float}
\usepackage{xcolor}

\newcommand{\yhyangCR}[1]{{\color{black}#1}}

\newcommand{\yhyang}[1]{#1}
\title{Melody Infilling with User-Provided Structural Context}

\multauthor
{Chih-Pin Tan$^{1,2}$  \hspace{1cm} Alvin W.Y. Su$^{1}$ \hspace{1cm} Yi-Hsuan Yang$^{2, 3}$}
{\\
$^1$~National Cheng Kung University,
$^2$~Academia Sinica,
$^3$~Taiwan AI Labs \\
{\tt\small p76091551@gs.ncku.edu.tw, alvinsu@mail.ncku.edu.tw, yang@citi.sinica.edu.tw}
}

\def\authorname{Chih-Pin Tan, Alvin W.Y. Su, and Yi-Hsuan Yang}

\usepackage[bookmarks=false,pdfauthor={\authorname},pdfsubject={\papersubject},hidelinks]{hyperref}
\begin{document}

\maketitle
\begin{abstract}
This paper proposes a novel Transformer-based model for music score infilling,  to generate a music passage that fills in the gap between given past and future contexts.
%While existing prompt-based conditioning approaches can
While existing infilling approaches can
generate a passage that connects smoothly locally with the given contexts, 
%Known researches have demonstrated that prompt-based conditioning approaches could make great results in local smoothness among past context, future context and the generated sequence.
%However, these 
they do not take into account the musical form or structure of the music  and may therefore generate overly smooth results.
%cannot guarantee the structures of musical context. 
%the repeatness and similarity corresponding to
To address this issue, we propose a structure-aware conditioning approach that employs a novel attention-selecting module to supply user-provided structure-related information to the Transformer for infilling.
With both objective and subjective evaluations, we show that the proposed model can harness the structural information effectively and generate melodies in the style of pop of higher quality than the two existing structure-agnostic infilling models.
%, including comparison of melody, rhythm and tonality, and human listening tests,
%The results show that our approach achieves a great improvement on generating pop music by efficiently taking advantage of music structure information.
\end{abstract}
\section{Introduction}\label{sec:introduction}
In recent years, machine learning techniques have been widely applied to symbolic music generation.
A large number of models attain \emph{sequential generation} by accounting for only the \textit{past context}, i.e., the generated music depends on only the preceding musical content \cite{boulanger2012modeling,colombo2016algorithmic,sturm2016music,roberts2018hierarchical,yang2017midinet,huang2018music,donahue2019lakhnes,huang2020pop,ren2020popmag,wu2020jazz,dai21ismir,musemorphose,symphonynet22arxiv,mmtransformer22arxiv}.
While sequential generation can find useful use cases, it does not align with typical human compositional practices which can be non-sequential in nature. Musicians often write motifs or small pieces to get inspiration first, before working on the middle parts to connect them.
%Sequential generative models accept past contexts only. Once future contexts are generated, there is no way to tweak them to satisfy the compositional requirements.
%Since sequential generative models consider only the past contexts, it is hard to tweak them to satisfy the compositional requirements where the \textit{future context}, the succeeding musical content, has to be incorporated into the generation process.

\begin{figure}[t]
 \centerline{
 \includegraphics[width=.95\columnwidth]{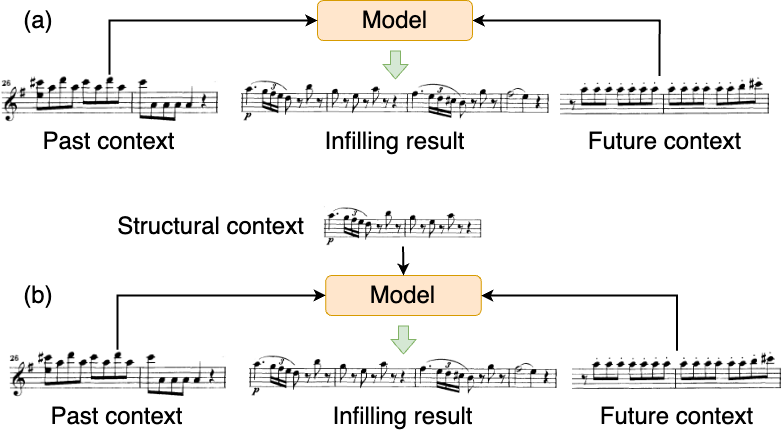}}
 \caption{Comparison between (a) structure-agnostic  and (b) structure-aware approaches for music score infilling.}
 \label{fig:overall}
\end{figure}

Hence, we focus on the scenario when both the past and \emph{future} contexts are given, which is called \textit{music score infilling} or inpainting \cite{pati2019learning}. 
As shown in \figref{fig:overall}(a), the task is to let models fill in the missing part
%missing information 
between the two given  segments.
%by taking into account the complete contexts, 
Prompt-based conditioning approaches \cite{huang2017counterpoint,hadjeres2017deepbach,ippolito2018infilling,pati2019learning,bazin2019nonoto,chang2021variable,hsu2021generating,tan21ismirLBD,wei2022music} have been applied to such a task in recent years, treating  the two given segments as the ``prompt.''
%This can be done, taking the recent variable-length infilling (VLI) model \cite{chang2021variable} as an example, by associating the past and future contexts with suitable positional encodings and then using a self-attention based model such as the XLNet \cite{yang2019xlnet} to generate the missing part 
Among them, the variable-length infilling model (VLI)  \cite{chang2021variable}  obtains promising results by adding special positional encodings to XLNet \cite{yang2019xlnet}, a permutation-based language model that is naturally suitable for generative tasks with given bi-directional contexts.
The experiment of VLI shows that their model is capable of connecting the past and future contexts smoothly locally for infilling solo piano passages of up to 4 bars (measures). 

Considering composers usually write musical pieces in a hierarchical manner \cite{lerdahl1996generative}, we note that prompt-based conditioning approaches have a strong limitation: they generate results with only consideration of local smoothness among the past context, future context, and result, without taking care of the overall musical \emph{form} or \emph{structure} of the music.
For instance, a composer may like to write a song in a musical form of \texttt{ABA'B'}. If we consider {the concatenation of the segments corresponding to \texttt{A} and \texttt{B} (i.e., \texttt{AB})} as the past context, and {the segment corresponding to} \texttt{B'} as the future context, and feed them to an existing infilling model, the model may generate a sequence that consists of similar melody and chord progression as {the segments corresponding to} \texttt{B} and \texttt{B'}, not the intended repetition or variation of {the segment corresponding to} \texttt{A}.
%We refer musical structure to the musical form while it is usually used as the song structure in pop music, the chord-related structure in jazz or some other musical terms. 

To address this issue, we propose in this paper a novel \textbf{structure-aware} setting for music infilling.
% conditioning approach that uses a sequence-to-sequence (seq2seq) Transformer model \cite{vaswani2017attention,dai2019transformer} to achieve long-term, structure-involved music score infilling.
As shown in \figref{fig:overall}(b),
besides the past and future contexts exploited by  conventional structure-agnostic, prompt-based models, 
out approach additionally capitalizes for the infilling task the \emph{structural context}, a music segment corresponding to a certain part of the whole music {that is supposed to share the same \emph{structure label} (such as \texttt{A} or \texttt{B}) with the missing segment.
Accordingly, besides local smoothness, the model also needs to consider the similarity between the infilled segment and the structural context.}
Here, we assume  the structural context is provided by a user, not generated by a model.
%While the prompt around the infilled target is relevant to structure, we consider that the user-specified structural context gives the model a sight to the overall musical structure.  
%The experiment shows that our model can generate up to 16-bar-length monophonic infilled sequences, and dramatically outperforms the baselines on considering the overall structural context.
For example, the user may {designate the segment corresponding to} \texttt{A} as the structural context, thereby inform the model with the intended musical form.

We improve upon the VLI model \cite{chang2021variable} in the following ways to realize structure-aware infilling.
First, we use the classic Transformer
\cite{vaswani2017attention,dai2019transformer,radford2018improving} instead of the more sophisticated XLNet \cite{yang2019xlnet} as the model backbone, to make it easier to add 
{a conditioning module}
% conditioning modules
to exploit the structural context. 
To improve the capability of the Transformer to account for bi-directional contexts, we propose two novel components, the \emph{bar-count-down technique} (Section \ref{sec:method:countdown}) and \emph{order embeddings} (Section \ref{sec:method:order}), which respectively give the model an explicit control of the length of the generated music, and a convenient way to attend to the future context. % the via self-attention. 
Second, being inspired by the Theme Transformer \cite{shih2022theme}, we use not a Transformer decoder-only architecture but a
%seq-to-seq
{sequence-to-sequence (seq2seq)} Transformer encoder/decoder architecture, using the cross-attention between the encoder and decoder as
%a
{the} conditioning module to account for the structural context.
Moreover, we propose an \emph{attention-selecting module} that allows the Transformer to access multiple structural contexts while infilling different parts of a music piece, which can be useful both in the training and inference time (Section \ref{sec:method:attn}) .

% While the design of our seq2seq Transformer is inspired by the Theme Transformer \cite{shih2022theme}, which is for sequential generation, our model involves three novel technical elements to achieve infilling.
% First, %in view that the vanilla Transformer is limited to sequential generation, 
% we propose a new positional segment embedding to ``reorder'' the given prompt sequences without breaking the original positional relationship, so that we can use the Transformer instead of the more sophisticated XLNet \cite{yang2019xlnet,chang2021variable} as our model backbone  (Section \ref{sec:method:order}).
% Second, we use a bar-count-down technique to control the number of infilled bars  (Section \ref{sec:method:countdown}).
% Third, we propose an attention-selecting module that allows the Transformer to access multiple structural contexts while infilling different parts of a music piece (Section \ref{sec:method:attn}).

%The experiment shows that our model can generate 
For evaluation, we compare our model with two strong baselines, the VLI \cite{chang2021variable} and the work of Hsu\,\&\,Chang \cite{hsu2021generating},
%, which are adapted for our token representation in the experiment.
on the task of symbolic-domain melody infilling \yhyangCR{of 4-bar content}
%with single user-provided structural context.
using the POP909 dataset  \cite{pop909-ismir2020} \yhyangCR{and the associated structural labels from Dai \emph{et al.} \cite{dai2020automatic}}.
%by transforming the music into the acceptable representations of their models.
With objective and subjective analyses, we show that our model greatly outperforms the baselines in the structure completeness of the generated pieces, without degrading local smoothness.

We set up a webpage for demos\footnote{\mbox{https://tanchihpin0517.github.io/structure-aware\_infilling}} and open source our code at a public GitHub repository.\footnote{\mbox{https://github.com/tanchihpin0517/structure-aware\_infilling}}

\begin{figure*}[t]
 \centerline{
 \includegraphics[width=1.8\columnwidth]{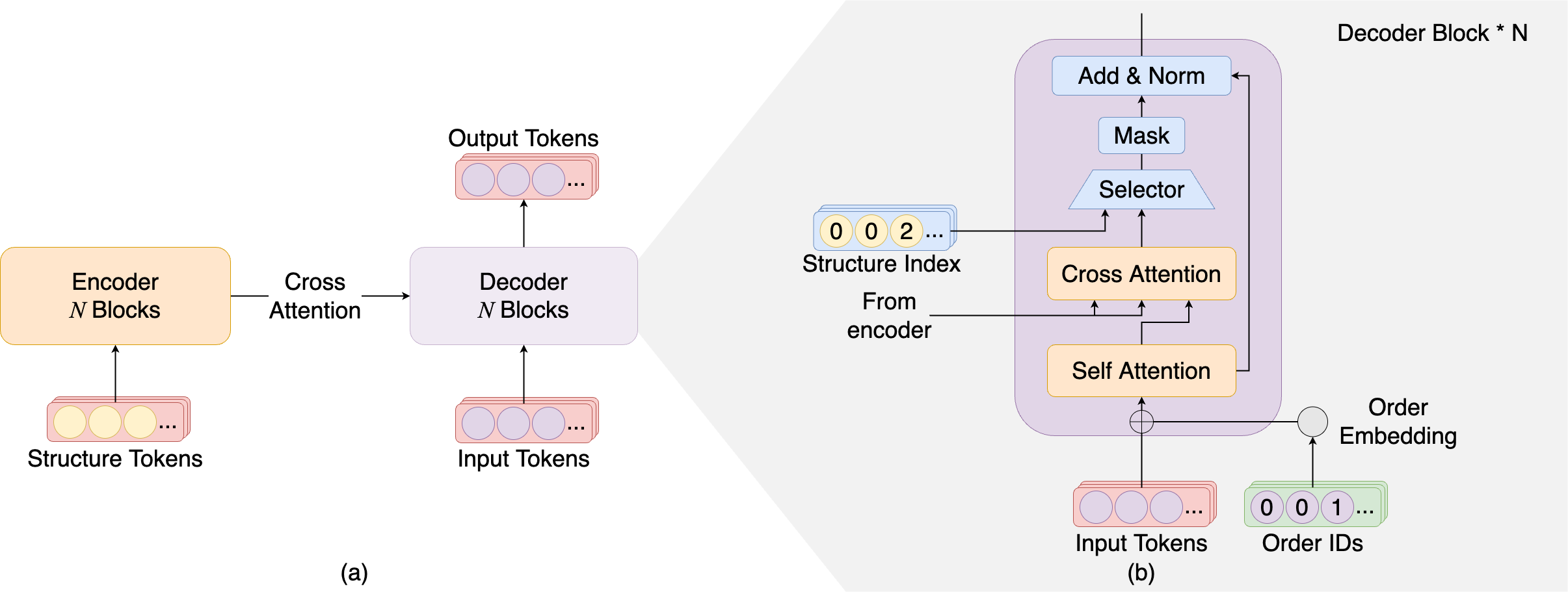}}
 \caption{Schematic diagram of (a) the proposed seq2seq Transformer model for structure-aware infilling, and (b) a zoom-in of the decoder highlighting how the decoder utilizes the \textit{order embedding} and \textit{structure index}.}
 \label{fig:arch}
\end{figure*}

\section{Related Work}\label{sec:related_work}

%\subsection{Infilling \& Transition}
% fixed length
%Deep Bach
Generating missing parts with given surrounding contexts has been attempted by early works.
DeepBach \cite{hadjeres2017deepbach} predicts missing notes based on the notes around them. They use two recurrent neural networks (RNNs) to capture the past and future contexts, and a feedforward neural network to capture the current context from notes with the same temporal position as the target note.
%Counterpoint by Convolution
COCONET \cite{huang2017counterpoint} trains a convolutional neural network (CNN) to complete partial musical scores and explores the use of blocked Gibbs sampling as an analog to rewriting. They encode the music data with the piano roll representation and treat that as a fixed-size image, \yhyangCR{so the model can only perform fixed-length music infilling}.
%Learning to traverse latent spaces for musical score inpainting
Inpainting Net \cite{pati2019learning} uses an RNN to integrate the temporal information from a variational auto-encoder (VAE) \cite{kingma2013auto} for bar-wise generation,
% Music Phrase Inpainting Using Long-Term Representation and Contrastive Loss
Wei \emph{et al.} \cite{wei2022music} build the model with a similar concept as Inpainting Net and use the contrastive loss \cite{chen2020simple,he2020momentum} for training to improve the infilling quality.
Some Transformer-based models have also been proposed to achieve music infilling.
% Infilling Piano Performances
Ippolito \emph{et al.} \cite{ippolito2018infilling} concatenate the past and future context with a special separator token. They keep the original positional encoding of the contexts and the missing segment, which \yhyangCR{again limits} the length of given contexts and generated sequence \yhyangCR{to be fixed}.
\yhyangCR{We see that these} infilling models impose some data assumptions and thereby have certain restrictions, e.g., the length of the input sequence cannot be arbitrary, or the missing segment needs to be complete bars. 
%such as the fixed-length input sequence and bar-wise start/end position of generation.
%Generating Music Transition by Using a Transformer-Based Model
The work of Hsu\,\&\,Chang \cite{hsu2021generating} is free of these restrictions. They use two Transformer encoders to capture the past and future context respectively and generate results with a Transformer decoder.
\yhyangCR{The VLI model \cite{chang2021variable} can also realize variable-length infilling.}
%In the experiment, we show that prompt-based models cannot capture structural context due to their property.
\yhyangCR{However, to our best knowledge,} no existing models have explicitly considered structure-related information for infilling.

Structure-based conditioning has been explored only recently by Shi \emph{et al}. \cite{shih2022theme}  in their Theme Transformer model for sequential music generation. They use a seq2seq Transformer to account for not only the past context but also an additional pre-given theme segment that is supposed to manifest itself multiple times in the model's generation result.
The present work can be considered as an extension of their work to the scenario of music infilling.

%\subsection{Structure-based Conditioning}
%
%Generative systems that use musical contexts or patterns to control the generating process have been explored in early works. For rule-based systems, Shan and Chiu \cite{shan2010algorithmic} generate a new music object based on the rules and patterns that are discovered by their data mining system.  Zalkow \emph{et al.} \cite{se2016musical} change the style of an existing musical work via their rule-based system. For deep learning works, Theme Transformer \cite{shih2022theme} uses a Transformer-based model with gated parallel attention to generate results from prompts and an additional theme, a given music segment. However, to our best knowledge, no existing work which explicitly employs music structural contexts or patterns for infilling tasks has been proposed before.

\section{Methodology}\label{sec:methodology}

%Following \cite{pati2019learning}, w
Given a past context $C_\text{past}$ and a future context $C_\text{future}$, the general, \textbf{structure-agnostic music infilling} task entails generating an \emph{infilled segment} $T$ that interconnects $C_\text{past}$ and $C_\text{future}$ \yhyangCR{smoothly, preferably} in a musically meaningful way.
%In other words, the training object of a general infilling model is to maximize the likelihood:
When using an autoregressive generative model such as the Transformer as the model backbone, the training object is to maximize the following likelihood function:
\begin{equation}
\label{eq:formulation:general}
    \prod_{0<k \leq |T|} P(t_k| t_{<k},C_\text{past},C_\text{future}) \,,
    %P(T|C_\text{past},C_\text{future}) \,.
\end{equation}
where $t_k$ denotes the element of $T$ at timestep $k$,  $t_{<k}$ the subsequence consisting of all the previously generated elements, and $|\cdot|$ the length of a sequence.

Extending from \eqnref{eq:formulation:general}, we propose and study in this paper a special case, called \textbf{structure-aware music infilling},
%where an additional segment $G$ representing the \emph{structural context} is given, leading to the new objective:
where an additional 
segment $G$ representing the
%group of segments $G$ representing the
\emph{structural context} is given, leading to the new objective:
%the training object can be factorized to:
\begin{equation}
\label{eq:formulation:struct}
    \prod_{0<k \leq |T|} P(t_k|t_{<k},C_\text{past},C_\text{future};G) \,.
\end{equation}
%We consider the case in this paper where our model is based on a seq2seq Transformer, which predicts $T$ autoregressively, and the model accepts extra user-provided structure information $C_s$.
%We consider the case of structure-based conditioning in this paper where our model is based on a seq2seq Transformer, which is of an encoder-decoder architecture in \figref{fig:arch}.
%We consider the case of structure-based conditioning where the structural context is given.
%Its input is comprised of two token sequences, the \textit{prompt context} \{$C_\text{past}$, $C_\text{future}$, a masked version of $T$\} and the \textit{structural context} \{$C_s$\}, that are the input tokens of decoder and encoder respectively. The training loss is computed over $T$ only.
As depicted in \figref{fig:arch}(a), our model is based on Transformer with the encoder-decoder architecture.
It uses the decoder to self-attend to the prompt (i.e., $C_\text{past}$ and $C_\text{future}$) and the previously-generated elements (i.e., $t_{<k}$), and the encoder to cross-attend to the structural context $G$. %Structure group is composed of several structural contexts which are user-provided music segments.
%The training loss is computed over $T$ only.
%Since Transformer is an autoregressive generative model\cite{radford2018improving} which generate sequences progressively, 
We provide details of the proposed model below.

Note that we do not require the length of all the involved segments to be fixed; namely $|T|$, $|C_\text{past}|$, $|C_\text{future}|$ and %$|\text{sequences in }G|$ 
$|G|$ are all variables in our setting.

\subsection{REMI-based Token Representation}
\begin{figure}[t]
 \centerline{
 \includegraphics[width=0.8\columnwidth]{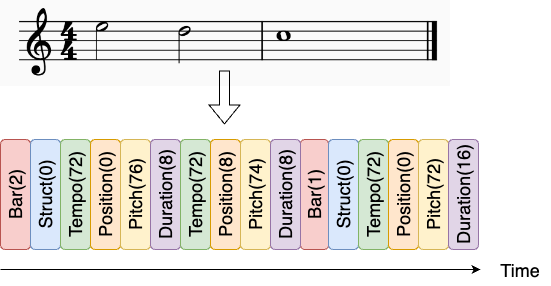}}
 \caption{An example of representing a two-bar  segment with our \yhyangCR{adapted} REMI-based token representation. With the bar-count-down technique, \texttt{Bar(2)} and \texttt{Bar(1)} are used for the first and second \texttt{Bar} tokens, respectively.}
 \label{fig:remi}
\end{figure}

{To incorporate structure-related information to our representation of the music data, we devise an extension of} the REMI-based representation \cite{huang2020pop} {that comprises}
%All musical content are translated to 
five types of tokens: \texttt{Bar}, \texttt{Struct}, \texttt{Tempo}, \texttt{Position}, \texttt{Pitch} and \texttt{Duration}.
\tabref{tab:vocabulary} \yhyangCR{lists} the vocabulary of our token representation.
\texttt{Bar} consists of numbers from 1 to 32, standing for the number of remaining bars on the generation process.
%We will explain the details of \texttt{Bar} in the next subsection.
A music form, e.g., \texttt{ABA'B'}, consists of multiple groups of similar {phrases or sections (each of multiple bars)}, e.g., \{\texttt{A},\,\texttt{A'}\} and \{\texttt{B},\,\texttt{B'}\}, where each group {can be said to be associated with the same \emph{structure label}}.
%the notes of a group are regraded as having the same structural context. 
%To include the information with the token representation, 
We use
\texttt{Struct} to {indicate the structure label for each bar.}
%represent the corresponding structural context of the current bar, where we assume the structural context is aligned to bars.}
\texttt{Tempo} and \texttt{Position} are related to the musical metre.
\texttt{Tempo} is the current tempo of beats per minute (BPM), and
\texttt{Position} is the temporal distance between the onset of a musical note and the beginning of its bar, denoted by the number of 16-th notes.
\texttt{Pitch} and \texttt{Duration} are related to musical notes, which are the MIDI pitch number and duration in 16-th notes, respectively.
We show an example of how we encode the musical content in \figref{fig:remi}.

\begin{table}[t]
 \begin{center}
 \begin{tabular}{l|c|l}
  \hline
  Token type & Voc. size & Values \\
  \hline
  \texttt{Bar}  & 32 & 1, 2, ... , 32 \\
  \texttt{Struct} & 16 & 0, 1, ... , 15 \\
  \texttt{Tempo}  & 47 & 28, 32, ... , 212 \\
  \texttt{Position}  & 16 & 0, 1, ... , 15 \\
  \texttt{Pitch} & 86 & 22, 23, ... , 107 \\
  \texttt{Duration}  & 16 & 1, 2, ... , 16 \\
  \hline
 \end{tabular}
\end{center}
 \caption{The vocabulary of the token representation.}% \texttt{Tempo} is in the unit of Beat-Per-Minutes (BPM).}% Note that special tokens such as \texttt{BOS} and \texttt{PAD} are not contained in this table since they are model-dependent.}
 \label{tab:vocabulary}
\end{table}

\subsection{Bar-Count-Down Technique}
\label{sec:method:countdown}

The work of Hsu\,\&\,Chang \cite{hsu2021generating} uses special \texttt{BOS} and \texttt{EOS} tokens as the ``signal'' to start or stop the generation process.
{At inference time, the infilled segment generated by their model comes to an end when the model generates an \texttt{EOS} token.
While this may work fine in certain cases, doing so cannot give us an explicit control of the number of bars to be generated for the infilled segment.
Such a control is preferable when we want to make sure that the previous context and the future context are a certain number of bars apart, which is highly needed for structure-aware infilling. For example, a user may want to specify the music form as \texttt{A8B8A'8B'8}, meaning that all the four sections \texttt{A}, \texttt{B}, \texttt{A'}, \texttt{B'} are eight-bar long each.}

%In order to control how many bars the model generates,
{To have such a control},
we employ a special token representation technique called ``bar-count-down.''
For each infilled sequence $T$, we adjust the suffix number of \texttt{Bar} tokens to match the length of $T$. Take \figref{fig:remi} as an example: The number in \texttt{Bar} tokens are counted down from two because the sequence in \figref{fig:remi} is 2-bars long.
Once training a model with this setup,
%we just need to give the first \texttt{Bar(N)} token to control the length of the infilled sequence on generation.
the length of $T$ can be controlled {effectively} by the {number of remaining bars associated with the} first \texttt{Bar} token given on generation.

\begin{figure}[t]
 \centerline{
 \includegraphics[width=0.9\columnwidth]{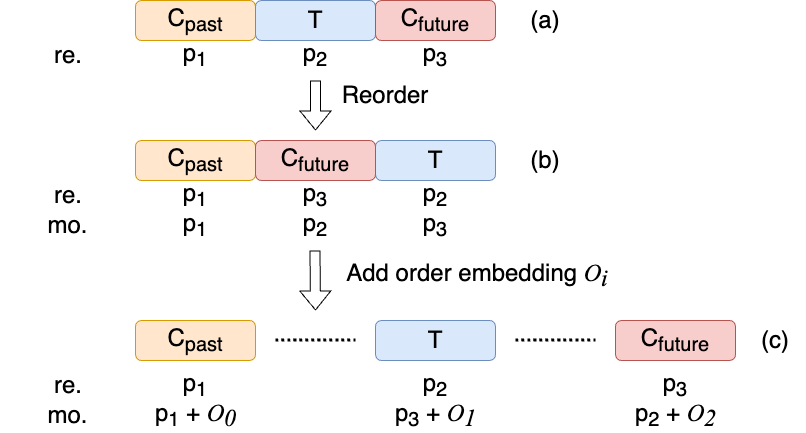}}
 \caption{Illustration of the order embedding, where $re.$ and $mo.$ stand for the \textit{real position} and \textit{model-viewed position}.
 %The \textit{real position} is the position in musical meaning, and the \textit{model position} is the position which the model views on computing attention scores. 
 \yhyangCR{Conceptually}, we shift the model-viewed position by $O_i$ to make it consistent with the real position.}
 \label{fig:reorder}
\end{figure}

\subsection{Order Embedding}
\label{sec:method:order}

Transformers with causal masking are mainly designed for sequential generation. To apply the model to infilling tasks where the missing part is in the middle of the input sequence, 
{the model proposed by Hsu\,\&\,Chang \cite{hsu2021generating} attends to bi-directional context} from two encoders with cross-attention.
%We propose a new approach that exploits bi-directional information in the decoder via only self-attention. Specifically, we 
{As we want to instead} use only the self-attention of the decoder to exploit bi-directional information, we reorder the  sequence \{$C_\text{past}$, $T$, $C_\text{future}$\} to \{$C_\text{past}$, $C_\text{future}$, $T$\}.
\yhyangCR{Doing so would however} change the original positional relationship among $C_\text{past}$, $T$, and $C_\text{future}$.\footnote{The principal idea of the attention mechanism \cite{vaswani2017attention} is to use the token embeddings of two tokens to compute their correlation, leading to the so-called attention score. However, using the token embeddings alone fails to consider the position-related relations of the two tokens (e.g., whether they are neighbors or distant apart). Accordingly, in practice, people add  token-wise positional embeddings to the token embeddings before computing their attention  \yhyangCR{\cite{shawSelfAttentionRelativePosition2018,keRethinkingPositionalEncoding2020,pmlr-v139-liutkus21a}}. } 
As depicted in \figref{fig:reorder}(b), 
%the \textit{real position} and \textit{model-viewed position}, which are the positional relationships of the tokens before and after reordering, are mismatched.
%As depicted in \figref{fig:reorder}, the real position and the model-viewed position of the sequence \{$C_\text{past}$, $T$, $C_\text{future}$\} are mismatched after reordering.
the \emph{real} positional relationships among the segments entails associating the tokens in $T$ with positional embeddings corresponding to a set of positions $p_2$ that signifies the model the segment $T$ is after $C_\text{past}$ (i.e., $p_1$ $\prec$ $p_2$) and is before $C_\text{future}$ ($p_2$ $\prec$ $p_3$).\footnote{{We use $p \prec q$ to denote that any elements in the set $p$ is smaller, or ``temporally before,'' any elements in the set $q$.}} After reordering, however, using the typical way of computing the positional embeddings from left to right by the Transformers, $T$ would be assigned with positional embeddings corresponding to $p_3$, meaning that by the \emph{model-viewed} positional relationships, $T$ is after both $C_\text{past}$ and $C_\text{future}$. The real and the model-viewed are mismatched.

%Considering the linearity of the positional relationship, we hypothesize this mismatch can be eliminated by adding an ``offset'' to the original positional embedding.

We introduce a new position-related \emph{segment embedding} called ``order embedding'' to tackle this issue.
%\footnote{Please refer to \cite{vaswani2017attention} and \cite{dai2019transformer} for details of the attention mechanism.}
%In some tasks such as sentence-pairing in natural language processing, the \textit{segment embedding} is used to cover the insufficiency of the positional embedding for grouping tokens in the input sequence \cite{devlin2018bert}.
%Inspired by previous works, the order embedding is the position-related segment embedding to help model to figure out the relationship of the segments in the input sequence, i.e., $C_\text{past}$, $T$, and $C_\text{future}$.
Specifically, for the positional embeddings for all the tokens in $C_\text{past}$, we add to them the same additional embedding corresponding to a positional ``offset'' or ``order,'' denoted as $O_0$. We similarly incorporate $O_1$ and $O_2$ to the positional embeddings for the tokens in $T$ and $C_\text{future}$. As long as the ``offset'' is big enough, we can have $p_1+O_0  \prec p_3+O_1 \prec p_2+O_2$, ensuring that the real and model-viewed positional relationships are matched, as depicted in \figref{fig:reorder}(c).

We note that, as the same order embedding is added to the positional embeddings of all the tokens in a segment, the proposed idea works nicely regardless of whether the segment lengths $|C_\text{past}|$, $|T|$, $|C_\text{future}|$ are fixed or not.

\subsection{Attention-Selecting Module}
\label{sec:method:attn}

\begin{figure}
 \centerline{
 \includegraphics[width=1.0\columnwidth]{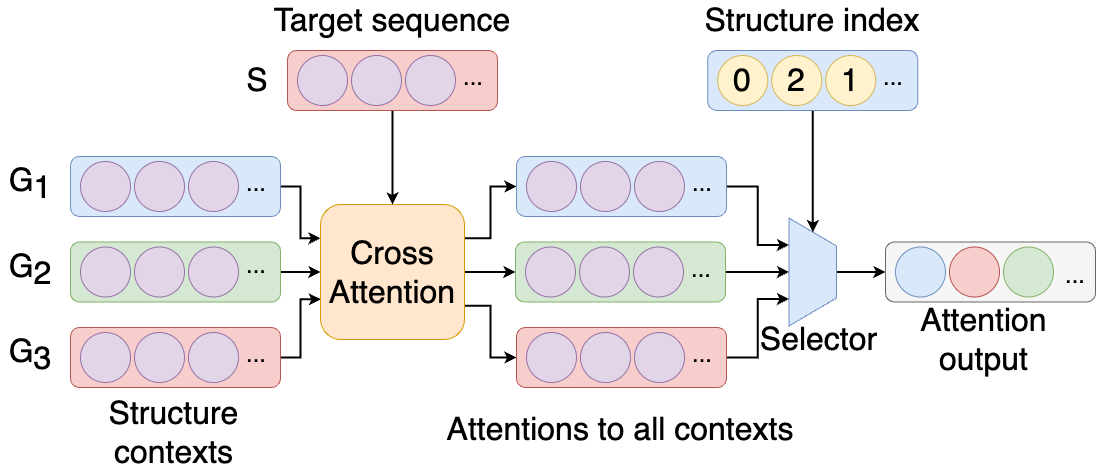}}
 \caption{The schematic of attention selecting. Multiple structural contexts $G_1,G_2,G_3$ are passed to the cross-attention block simultaneously, and the output is filtered by the \textit{selector} with the \textit{structure index}.}
 \label{fig:select}
\end{figure}

{While the formulation of \eqnref{eq:formulation:struct} considers only one structural context $G$, in practice, we may want to designate multiple structural contexts \{$G_1$, $G_2$, \dots, $G_N$\} for infilling, with each segment $G_n$ corresponding to a certain structure label such as \texttt{A} and \texttt{B}.} 
\yhyang{This is useful, for example, when the first part of the infilled segment $T$ is meant to be similar to phrase \texttt{A}, while the latter part similar to phrase \texttt{B}.
To indicate which $G_n$ a specific token $t_k$ of $T$ should refer to, we define the \emph{structure index} $y_k \in \{0,\dots,N\}$, for $k=\{1,\dots,|T|\}$. 
The structure indices are also given by the user when the user specifies the intended musical form. With all these,} we extend \eqnref{eq:formulation:struct} to:
\begin{equation}
\label{eq:formulation:struct_ext}
    \prod_{0<k \leq |T|} P(t_k|t_{<k},C_\text{past},C_\text{future};G_1,...,G_N, y_k) \,.
\end{equation}
%The model accepts structural contexts $G_n$
%%, which contains $N$ sequences $\{c_{1}$, $c_{2}$, ..., $c_{N}$\}, 
%to get the structural information of tokens in the input sequence. 
%%For a musical form, e.g., $A_1B_1A_2B_2$, each sequence in $G$ is a part of the structural context belonging to different similar phrases, e.g., $(A_1,A_2)$.
%Since \tanch{the tokens may belong to different structure labels}, the \eqnref{eq:formulation:struct} is extended to:
%For each $t_k$, it may belong to (i) a corresponding structural context $G_n$ or (ii) none of the structural contexts. The second case is that $t_k$ is in the phrases such as bridge, where there is no other similar phrase to them.
\yhyang{We use $y_k=0$ to indicate the case where $t_k$ is supposed to follow none of the structural contexts (e.g., when $t_k$ is part of the bridge). When $y_k\neq 0$, $t_k$ follows only one of the structural context $G_{y_k}$.
In our implementation, for tokens whose $y_k=0$, we only use the self-attention to attend to the prompt and  $t_{<k}$, leading to a formulation akin to \eqnref{eq:formulation:general}. When $y_k\neq 0$, we have a ``selector'' that picks by $y_k$ the structural context $G_{y_k}$ to be attended to via cross-attention, resulting in a formulation akin to \eqnref{eq:formulation:struct}. This is depicted in \figref{fig:arch}(b) and \figref{fig:select}.}

%We propose an attention-selecting mechanism to have tokens attend to the right structural context in the phase of \textit{cross-attention} in \figref{fig:arch}(b).
%We choose cross-attention since it outperforms other kinds of attention mechanisms in integrating external information \cite{shih2022theme}. 

%The structural context $C_s$ contains $N$ sequences \{$C_{s_1}$, $C_{s_2}$,...,$C_{s_N}$\} and each token in prompt sequences $S = \{s_1,...,s_L\} = \{C_\text{past}, T, C_\text{future}\}$ corresponds to structural contexts $\{c_1,../.c_L\}$, where $c_i \subseteq \{C_{s_1}$,...,$C_{s_N}\}$.

%Let the structure group $G$ contain $N$ structural contexts \{$C_{s_1}$, $C_{s_2}$, ..., $C_{s_N}$\} and 
%$\{c_1, ..., c_L\}$ be the structural contexts of each token in the prompt sequence $S = \{C_\text{past}, T, C_\text{future}\} = \{s_1, ..., s_L\}$ of length $L$, where $c_i \subseteq G$.
%Let the input sequence $S$ = $\{C_\text{past}, T, C_\text{future}\}$ which is composed by $L$ tokens $\{s_1, ..., s_L\}$.
%We define \textit{structure index} $I$ = $\{i_1,...,i_L\}$ with $i_t$ $\in$ $[1, L]$ which indicates that the token $s_t$ corresponds to the the structural context $G_{i_t}$.

\yhyang{\figref{fig:select} also shows that, in our implementation, instead of computing the cross attention between the target sequence with a specific structural context $G_{y_k}$, we actually compute the cross attention between the target sequence with every structural contexts  \{$G_1$, $G_2$, \dots, $G_N$\} and let the selector  pick the right one. While this seems a waste of computing, we find doing so faster when GPUs are used.}\footnote{\yhyang{There is another implementation detail: actually, not only the tokens in $T$ but also those in $C_\text{past}$ and $C_\text{future}$ would go through the Transformer's self- and cross-attention blocks. This is to get the latent vectors for the tokens in $C_\text{past}$ and $C_\text{future}$. In doing so, we calculate and employ the structure indices for $C_\text{past}$ and $C_\text{future}$ as well.}}

\section{Experimental Setup}
%\subsection{Dataset}
%\issue{mention the number of struct labels and adding BCD to hsu's model}.

We collect the melody data\footnote{Please note that the ``melody'' data in this paper is not exact monophonic music. We merge two midi tracks, ``MELODY'' and ``BRIDGE'' of the MIDI files of POP909 \cite{pop909-ismir2020}  to generate the melody data, which is monophonic most of the time but not always.} from the POP909 dataset compiled by Wang \emph{et al.} \cite{pop909-ismir2020}. POP909 contains 909 MIDI files of pop piano performances,
though we discard 8 of them  due to errors encountered in the preprocessing stage.
We split all remaining songs into 811 (90\%) songs for training and 91 (10\%) songs for testing.
The 16-th note is set as the minimal temporal resolution to quantize the tempo, beat, and duration for reducing the vocabulary size.

Dai \emph{et al.} \cite{dai2020automatic} publicly share the \emph{structural information} of songs in POP909 with letters and integers to indicate the structure labels and their lengths in bars,
%of the \issue{structural contexts
such as
\begin{equation}\label{eq:struct}
    \texttt{i4\,A8\,B8\,x4\,A8\,B8\,B8\,X2\,c4\,c4\,X2\,B9\,o2} \,,
\end{equation}
%The different letters represent the different repeated or similar phrases, and the phrases with the same letters are considered to share the same structural context.
where the music phrases labeled with the same letter are considered to share the same structural context.\footnote{The lower-case and capital letters indicate respectively  \emph{non-melodic} and \emph{melodic} phrases (i.e., where a clear melody is present, mostly a vocal line or an instrument solo), \yhyangCR{but we do not use this information}.}\footnote{Depending on the underlying musical form of a song, different songs may have different numbers of phrase groups.  For example, \yhyangCR{a song with a simpler form} may only have phrase groups \texttt{A} and \texttt{B}, while other songs have much more. In the POP909 dataset, a song can contain up to 9 unique phrase groups, which are labeled as \texttt{A}, \texttt{B}, \texttt{C}, \texttt{D}, etc.}
%different letters are different structural phrases and numbers are the bar-length of phrases.
Besides, the musical phrases with the labels \texttt{i}, \texttt{x}, and \texttt{o} are respectively the introduction, bridge, and ending, \yhyangCR{which do not have structural context in our setting}.
%We extract the structure corresponding to different letters from the first phrase labeled with the same letter.
For each song, we use the phrases corresponding to the first occurrence of the structure labels such as \texttt{A} and \texttt{B} as the structural contexts $G_1, G_2$, etc.
%For example, the first \texttt{A8} and \texttt{B8} in (\ref{eq:struct}) are used as the structural context for phrases with letters \texttt{A} and \texttt{B}, respectively.
For example, we choose bars 5--12 (i.e., the \texttt{A8} phrase after \texttt{i4}) and bars 13--20 (i.e., the \texttt{B8} phrase before \texttt{x4}) in the example shown in \eqnref{eq:struct} as the structural contexts for \texttt{A} and \texttt{B} since they are the first music phrases in the song with the corresponding labels.

% how we generate training and testing data
%811 songs for training, 91 songs for testing
%We split the data files for training and testing, which contain 90\% and 10\% of total, respectively.

The training data is generated with the following steps: 
%(i) iterating and setting all phrases as the infilled sequence $T$ except the first and last ones, and 
(i) iterating through all labels except for the first and last ones in the structural information,
(ii) choosing their corresponding music phrases as the infilled sequences $T$, and
(iii) concatenating $T$ with their preceding and following 6-bars music segments to get $\{C_\text{past}, T, C_\text{future}\}$, yielding 8,607 data in total.
Before feeding the training data into the model, we reorder the input sequences and insert additional special tokens, \texttt{BOS}, \texttt{SEP}, and \texttt{EOS}, to change them into $\{\texttt{BOS}, C_\text{past}, \texttt{SEP}, C_\text{future}, \texttt{SEP}, T, \texttt{EOS}\}$.
The reordered input sequence and structural contexts  are transformed to the embeddings with size 512.
The model consists of an encoder and a decoder, where both of them consist of six 8-head self-attention layers with intermediate layers of dimension 2,048. Each layer of the encoder and decoder is connected with an 8-head cross-attention, as shown in \figref{fig:arch}.
The output from the model is transformed back to a probability distribution of the vocabulary with the softmax function. At inference time, we use nucleus sampling \cite{holtzman2019curious} to sample the output tokens with the threshold value of 0.9.

%We set 16-th note as the minimal temporal resolution to quantize all the remaining songs and fit tokens into the vocabulary in \tabref{tab:vocabulary} with our representation.
%The structural contexts are extracted from the phrases which first time they are labeled with their structure, such as the first \texttt{A8} and \texttt{B8} in (\ref{eq:struct}).
%We iterate all structure phrases except the first and last ones, and combine each phrase with its preceding and following 6 bars to a token sequence. 
%8607 data for training.
%reorder the input sequence and add \texttt{SEP}

%156 songs

%Since the compared models generate reasonable results while the number of bars of the generated content equal or less than 16.
%For comparison with the baseline models, the testing data is fixed in the length of 16 bars.
%We compare our model with baseline models in short-phrase and long-phrase settings\footnote{We implement bar-count-down technique on Hsu's model to control generated length.}.
%We choose two structure patterns, \texttt{AAB} and \texttt{ABB}, from testing songs for two reasons.
%We design the experiment with the following concept in mind.
%To evaluate the effection of  We choose two
%We compare our model with baseline models in short-phrase and long-phrase settings\footnote{We implement bar-count-down technique on Hsu's model to control generated length.}.
For evaluation, we create the testing data by: (i) searching from the testing songs all the 4-bar phrases \yhyangCR{that correspond to only a structure label}, (ii) keeping only the phrases that share the same structure label with one of its two neighboring phrases but a different structure label with the other neighbor, and (iii) setting the 4-bar phrase as the target $T$ and concatenate them with their preceding and succeeding 6-bar music segments, which are set as the past context $C_\text{past}$ and future context $C_\text{future}$, respectively.
%In the experiment, we compare the models with the testing data fixed in the length of 16 bars\footnote{We implement bar-count-down technique on Hsu's model to control generated length.}.
%All music segments with two structural patterns, \texttt{AAB} and \texttt{ABB}, are collected from the testing songs and the segments of the middle label with 4-bars length are reserved.
%We set the middle segments as the targets and concatenate them with their preceding and succeeding 6 bars music segments to get the 16-bars prompt sequences.
%In the long-phrase setting, we pick up the \issue{label pairs} ${(p_1, p'_1), ..., (p_k, p'_k)}$, where (i) the length of $p_i$ and $p'_i$ of the pair $(p_i, p'_i)$ is 4 bars and (ii) the range $[p_i, p'_i]$ of the pair $(p_i, p'_i)$ is less than 32 bars, from all pairs of the same labels.
%Then we set the segments of each picked-up pairs as the targets and connect them with their preceding and succeeding 16 bars music segments to get the prompt sequences with length less than 64 bars.
We get 156 test cases \yhyangCR{of \{$C_\text{past}$, $T$, $C_\text{future}$\}, each with 16 bars (i.e., $6+4+6$)}. All the cases have the form of, e.g., \texttt{AA'B} or \texttt{ABB'}, and the target lengths are \yhyangCR{\textbf{4 bars} with arbitrary number of notes (hence \emph{variable} sequence lengths)}.
\yhyangCR{In this setting, the attention selection mechanism is used only for training, since the target sequence $T$ in our testing data only have one structural context to refer to.}

%target tokens of each testing data always have the same structural context (i.e., $\forall y_k$ $=$ $c$ where $c \neq 0$), the attention selection mechanism is used for training but not inference in the experiment.

We consider VLI \cite{chang2021variable} and the model from  Hsu\,\&\,Chang \cite{hsu2021generating} as the baselines.  
\yhyangCR{By design, only our model has access to} an external structural context.
\yhyangCR{However, we consider the comparison as valid}, since $C_\text{past}$ and $C_\text{future}$ are \yhyangCR{presumably} long enough to provide sufficient context, and at least one of them has the same structure label as $T$. 
Besides, in our implementation, we found the model  Hsu\,\&\,Chang \cite{hsu2021generating} rarely generates infilled segments with the desirable number of bars.  Therefore, \yhyangCR{we slightly improve their model by incorporating the bar-count-down technique}.
%Moreover, we design the experiment with the following conditions in mind.
%First, $C_\text{past}$ and $C_\text{future}$ should be long enough to provide sufficient contextual information.
%Second, one and only one of $C_\text{past}$ and $C_\text{future}$ shares the structural context with the infilled segment $T$.
%If neither does, the experiment becomes unfair because we provide the additional structural context to our structure-based model.
%If both do, early works have demonstrated that prompt-based models have the capability of connecting two similar music segments with the same pattern.
%Hence, we collect all music segments with two structural patterns, \texttt{AAB} and \texttt{ABB}, from the testing songs and
%examine whether the middle structural contexts of them are in the length of 4 bars. There are 156 music segments meeting the requirement.
%We concatenate them with their preceding and following 6 bars music segments to get the 16-bars prompt sequences for the experiment.%evaluations.

\begin{table}
 \begin{center}
 \begin{tabular}{lccc}
  \hline
    & $H \downarrow$ & $GS \uparrow$ & $D \downarrow$ \\
  \hline
    % Ours & \textbf{2.73$\pm0.72$} & \textbf{0.70$\pm0.08$} & \textbf{8.03$\pm12.70$}  \\ % note: the decimal should contain 2 digital even the number is zero
    % VLI\cite{chang2021variable} & 3.39$\pm1.56$ & 0.66$\pm0.10$ & 45.33$\pm25.81$  \\
    % Hsu \emph{et al.}\cite{hsu2021generating} & 9.99$\pm4.43$ & 0.65$\pm0.09$ & 63.96$\pm29.62$  \\
    % \hline
    % Real & 2.78$\pm0.89$ & 0.70$\pm0.09$ & 0.00$\pm0.00$  \\
    
    Ours & \textbf{2.75$\pm0.80$} & \textbf{0.70$\pm0.08$} & \textbf{25.73$\pm19.45$}  \\ % note: the decimal should contain 2 digital even the number is zero
    VLI\cite{chang2021variable} & 3.47$\pm1.57$ & 0.67$\pm0.09$ & 49.40$\pm25.12$  \\
    Hsu \cite{hsu2021generating} & 9.87$\pm4.64$ & 0.64$\pm0.09$ & 65.41$\pm38.00$  \\
    %\hline
    %Ours (long) & \textbf{3.64$\pm2.24$} & \textbf{0.65$\pm0.08$} & \textbf{105.79$\pm223.98$}  \\ % note: the decimal should contain 2 digital even the number is zero
    %VLI (long) & 3.07$\pm1.07$ & 0.64$\pm0.09$ & 46.13$\pm24.58$  \\
    %Hsu (long) & 9.79$\pm4.86$ & 0.61$\pm0.11$ & 45.18$\pm67.31$  \\
    % \hline
    % Copy & 2.77$\pm1.10$ & 0.70$\pm0.08$ & 22.46$\pm20.73$  \\
    \hline
    Original & 2.78$\pm0.89$ & 0.70$\pm0.09$ & ~~0.00$\pm~~0.00$  \\
  \hline
 \end{tabular}
\end{center}
 \caption{Objective evaluation results. 
 %in the short-phrase setting (1st row) and long-phrase setting (2nd row), 
 $H$: pitch class histogram cross entropy, $GS$: grooving pattern similarity, $D$: melody distance ($\uparrow$/$\downarrow$: the higher/lower the better).  %\yhyangCR{The `Copy' method is discussed in Section \ref{sec:subj}}.
 }
 \label{tab:hgsd}
\end{table}

% \begin{table}
%  \begin{center}
%  \begin{tabular}{lccc}
%   \hline
%     & $H \downarrow$ & $GS \uparrow$ & $D \downarrow$ \\
%   \hline
%     Ours & \textbf{2.75$\pm0.80$} & \textbf{0.70$\pm0.08$} & \textbf{25.73$\pm19.45$}  \\ % note: the decimal should contain 2 digital even the number is zero
%     Copy & 2.77$\pm1.10$ & 0.70$\pm0.08$ & 22.46$\pm20.73$  \\
%     \hline
%     Original\cite{pop909-ismir2020} & 2.78$\pm0.89$ & 0.70$\pm0.09$ & 0.00$\pm0.00$  \\
%   \hline
%  \end{tabular}
% \end{center}
%  \caption{The evaluation of copying the structural contexts as the results. ($\uparrow$/$\downarrow$: the higher/lower the better)}
%  \label{tab:copy}
% \end{table}

\section{Objective Evaluation Results}
\label{sec:obj}
%\subsection{Objective Evaluation}

\yhyangCR{
We propose three new metrics for objective evaluation, all of which have not been used in the literature of music infilling.
The first two metrics, \textbf{pitch class histogram cross entropy} ($H$) and \textbf{grooving pattern similarity}  ($GS$), are extensions of the ones proposed by Wu\,\&\,Yang for sequential generation \cite{wu2020jazz}  to our infilling task, evaluating respectively the consistency in terms of pitch class distribution (which is related to tonality) and rhythmic pattern. For $H$, we compute per test case the pitch class histogram of $T$, and 
that histogram of the concatenation of $C_\text{past}$ and $C_\text{future}$, and then report the cross entropy between these two histograms. For $GS$, we use per bar a 16-dim binary vector indicating where there is at least a note onset for every position in a bar, calculate one minus the normalized XOR difference between every pair of bars \cite{wu2020jazz}, one from $T$ and the other from either $C_\text{past}$ and $C_\text{future}$, and then report the average per test case.
The third metric, \textbf{melody distance} ($D$) measures the melody distance (dissimilarity) between
the infilled segment $T$ and the ground truth one (denoted as $T\#$) using the algorithm proposed by Hu \emph{et al.} \cite{hu2002probabilistic}.
Lower $H$ and higher $GS$ may imply that $T$ connects $C_\text{past}$ and $C_\text{future}$ smoothly, while lower $D$ indicates that the generation result is similar to a human-made one.
}

%We evaluate all the models with three metrics, \textit{pitch class histogram entropy} \cite{wu2020jazz}, \textit{grooving pattern similarity} \cite{wu2020jazz}, and \textit{melody distance} \cite{hu2002probabilistic}.
%The first one provides us an indicator of the music quality of the tonality of two music segments ($H$), the second one evaluates the rhythmic pattern similarity of two music segments ($GS$), and the last one indicates the melody distance (dissimilarity) between two music segments ($D$).
%$H$ and $GS$ are calculated on the pair $(T, \{C_\text{past}, C_\text{future}\})$,\footnote{$H$ and $GS$ are proposed as the bar-wise metrics in the original paper. In the experiment, we calculate $H$ over the full target sequences and $GS$ with the mean from all pairs of the bar of the target sequences.}
%and $D$ is calculated on the pair $(T, T\#)$, where $T\#$ denotes the ground truth of the target sequence from the real data.
%Since the goal of the task is to connect past and future contexts fluently and preserve the structural information in the generated music, the better model should have a lower value of $H$ and $D$, and a higher value of $GS$.
%Please note that $D$ doesn't include the ground truth because it's always zero.

\begin{figure*}[ht]
 \centerline{
 \includegraphics[width=2.0\columnwidth]{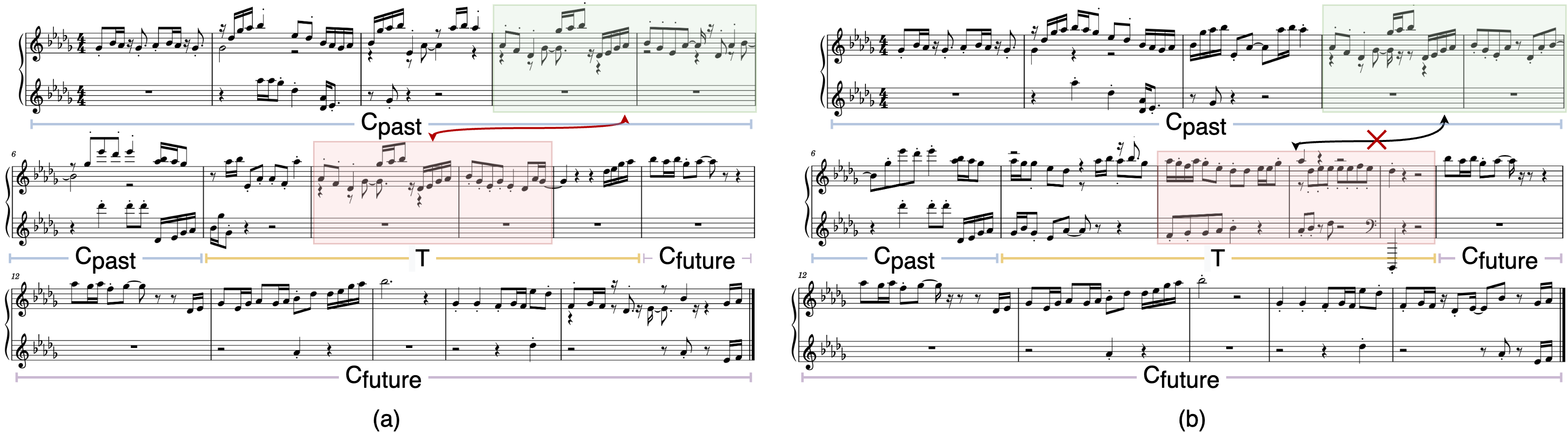}}
 \caption{Results generated by (a) our model and (b) VLI. The structural relations of bars 4-5 (green area) and bars 8-9 (red area) are preserved well in (a) but not in (b).}
 \label{fig:ex1}
\end{figure*}

\yhyangCR{
\tabref{tab:hgsd} shows that
%in the short-phrase setting, 
our model achieves the best result in all the three metrics, followed by VLI and then the model of Hsu\,\&\,Chang. 
Besides being consistent with the contexts, the infilling result of our model is closest to the ground truth one $T\#$, demonstrating the effectiveness of exploiting the structural context. 
\figref{fig:ex1} exemplifies how the infilled bars by our model fit the desired musical form.
}

\begin{table}
 \begin{center}
 \begin{tabular}{c|l|ccc|c}
  \hline
    \multicolumn{2}{c|}{} & M & R & S & O \\
  \hline
    \multirow{4}{1em}{all} & Ours & \textbf{3.46} & \textbf{3.51} & \textbf{3.40} & \textbf{3.42}  \\ % note: the decimal should contain 2 digital even the number is zero
    & VLI\cite{chang2021variable} & 2.96 & 3.14 & 3.12 & 2.97  \\
    & Hsu \cite{hsu2021generating} & 2.60 & 2.95 & 2.75 & 2.64  \\
    & Real & \textbf{3.77} & \textbf{3.77} & \textbf{3.62} & \textbf{3.66}  \\
  \hline
    \multirow{4}{1em}{pro}  & Ours & \textbf{3.58} & \textbf{3.28} & \textbf{3.28} & \textbf{3.42}  \\
    & VLI\cite{chang2021variable} & 2.67 & 2.86 & 2.78 & 2.72  \\
    & Hsu \cite{hsu2021generating} & 2.36 & 2.75 & 2.39 & 2.44  \\
    & Real & \textbf{3.61} & \textbf{3.56} & \textbf{3.42} & \textbf{3.42}  \\
  \hline
 \end{tabular}
\end{center}
 \caption{Results of the user study: mean opinion scores in 1--5 in \textbf{M} (melodic fluency), \textbf{R} (rhythmic fluency), \textbf{S} (Structureness), and \textbf{O} (Overall), from 'all' the participants or only the music 'pro'-fessionals.}
 \label{tab:user}
\end{table}

\section{Subjective Evaluation Results}
\label{sec:subj}
%\subsection{Subjective Evaluation}
\yhyangCR{We conduct additionally an online user study for subjective evaluation.
We have 91 anonymous volunteers}, where 12 of them are marked as professionals according to the question about their musical background.
Each subject is presented with 3 out of 15 sets of music segments randomly sampled from the testing data.  We inform them that the first and last 6 bars are the given prompts, and the middle 4 bars are the music generated by a model.
Each set of music contains \yhyangCR{in random order} 4 music including 3 generated by the models and 1 from the real data.
The subjects rate each of the 4 music 
\yhyangCR{in a 5-point Likert scale (the higher the better)}
according to their
(i) \textbf{melodic fluency}: do the pitches of notes go in the right tonality and connect the contexts fluently? 
%The more reasonable, the higher the score;
(ii) \textbf{rhythmic fluency}: are the notes played on the right beats? 
%The higher the accuracy rate, the higher score;
(iii) \textbf{structureness}: how is the generated part of the music similar to its contexts? 
%The more similar, the higher score;
(ix) \textbf{overall}: how much do they like the music? 
%The more they appreciate it, the higher their score.

\tabref{tab:user} shows the mean opinion scores (MOS) of the user study.
\yhyangCR{Echoing the result of the objective evaluation,} the proposed model outscores the baselines by a large margin \yhyangCR{in all the four subjective metrics}, and is close to the real music with a small gap.
\yhyangCR{The same observations can be made from either the average result of all the subjects, or only that from the 12 professionals.}

\begin{figure}[]
 \centerline{
 \includegraphics[width=1.0\columnwidth]{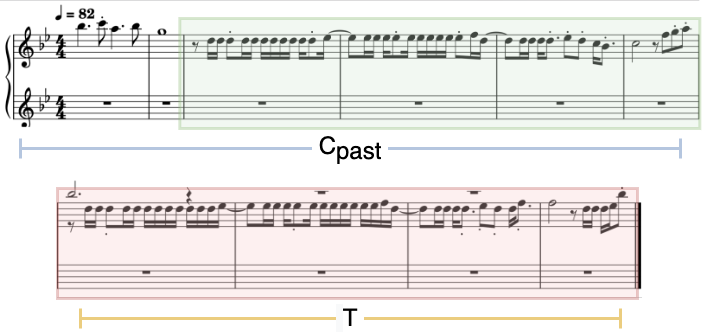}}
 \caption{A result generated by our model, where $C_\text{future}$ is not shown in this figure. Bars 3--6 (green area) and bars 7--10 (red area) look identical since the model overly imitates the given structural context. \yhyangCR{(Best viewed in color.)}}
 \label{fig:ex2}
\end{figure}

However, we notice that our model \yhyangCR{may} overly imitate the given structural contexts in some cases, as exemplified in \figref{fig:ex2}. \yhyangCR{When this happens}, the generated music sounds rigid and non-creative.
\yhyangCR{We conjecture that this can be attributed to the limited diversity of} 
our training data---the \yhyangCR{melodies} corresponding to the same structure label in POP909 appear to be too similar to each other, which \yhyangCR{may not be uncommon for pop songs}.
% However, \yhyangCR{the additional `Copy' baseline
% shown in \tabref{tab:hgsd}, which simply} copies the structural contexts as the result, \yhyangCR{shows that our model is beyond simply copying.}
% While `Copy' gains high scores, our model \yhyangCR{obtains slightly higher $D$, implying that the generated results have a certain degree of diversity.}
To study this, we implement additionally a `Copy' baseline that simply copies the structural contexts as the result, and include its infilling result to the demo website.
Our own subjective listening of its result confirms that the proposed method still outperforms the `Copy' baseline most of the time,
as the connections between the targets and their contexts are considered by our model, but not by the `Copy' baseline.
% Moreover,  ‘Copy’ baseline overly imitate

%From our observation, the reason our model performs so great is that the pop music contains a lot of repeated melodies and chord progression, and our model can generate the infilled results in high quality by imitating the given structural context.

\section{Conclusion}
In this paper, we have proposed a new structure-aware conditioning approach for music score infilling.
To help Transformers exploit bi-directional contexts, we employ the order embedding to shift the position viewed by the model.
Besides, we introduce a new attention-selecting mechanism to account for multiple structural contexts. 
Evaluations \yhyangCR{on 4-bar melody infilling validate the superiority of the proposed model over two existing Transformer-based structure-agnostic infilling methods \cite{chang2021variable,hsu2021generating}}.

\yhyangCR{We can extend this work in three ways. First, instead of relying on user inputs, we may build}
%Nevertheless, the structural contexts are still user-specified in this work. It's worth considering how to make 
models that \yhyangCR{generate the musical form automatically and predict the structural context for a specific infilled segment to refer to.}
%and generate the corresponding infilled results fully automatically. 
\yhyangCR{Second, we like to expand our work to other music genres and polyphonic music.}
%Trying another dataset with songs in different music styles and improving our model to avoid generating songs with pure imitation is also important for future works. 
% By removing the order embedding, our model can be applied to sequential generation.
Finally, we like to study how the attention-selecting mechanism can be applied to sequential generative tasks 
\yhyangCR{such as theme-based generation \cite{shih2022theme}.}
%sequential generative tasks by removing the order embedding.

\newpage

\section{Acknowledgement}
We are grateful to Shih-Lun Wu for sharing the \yhyangCR{unpublished} idea of the bar-count-down technique,
Ping-Yi Chen and Chin-Jui Chang for the assistance with experiments and discussion,
Shan Lee for the help of figure drawing,
and Chun-Wei Lai and Sophia Lin for the help in finding volunteers for the user study.
We also thank the anonymous reviewers for their valuable feedbacks. Our research is funded by grants NSTC 109-2628-E-001-002-MY2 and NSTC 110-2221-E-006-137-MY3 from the National Science and Technology Council of Taiwan.

% For bibtex users:
%\newpage
\bibliography{reference}

% For non bibtex users:
%\begin{thebibliography}{citations}
% \bibitem{Author:17}
% E.~Author and B.~Authour, ``The title of the conference paper,'' in {\em Proc.
% of the Int. Society for Music Information Retrieval Conf.}, (Suzhou, China),
% pp.~111--117, 2017.
%
% \bibitem{Someone:10}
% A.~Someone, B.~Someone, and C.~Someone, ``The title of the journal paper,''
%  {\em Journal of New Music Research}, vol.~A, pp.~111--222, September 2010.
%
% \bibitem{Person:20}
% O.~Person, {\em Title of the Book}.
% \newblock Montr\'{e}al, Canada: McGill-Queen's University Press, 2021.
%
% \bibitem{Person:09}
% F.~Person and S.~Person, ``Title of a chapter this book,'' in {\em A Book
% Containing Delightful Chapters} (A.~G. Editor, ed.), pp.~58--102, Tokyo,
% Japan: The Publisher, 2009.
%
%
%\end{thebibliography}

\newpage

\end{document}